\documentstyle[12pt]{article}

\title{ANOMALY IN THE CHARGE RADII OF Pb ISOTOPES}
\author{M.M. SHARMA, G.A. LALAZISSIS and P. RING \\
Physik Department, Technische Universit\"at
M\"unchen \\
D-85747 Garching, Germany}

\begin{document}

\maketitle
\begin{abstract}
The anomalous behaviour of the charge radii of the isotopic chain of
Pb nuclei has been studied in the relativistic mean field theory.
It has been shown that the relativistic mean field
provides an excellent description of the anomalous kink in the isotopic
shifts about $^{208}$Pb. This contrasts strongly from the Skyrme mean
field, where the density-dependent Skyrme forces fail to reproduce
the observed trend in the empirical data on the charge radii. The results
have been discussed in the perspective of differences in the
ans\"atze of the relativistic and the Skyrme mean-field theories.
\end{abstract}

\newpage
\baselineskip 15.5 pt

The relativistic mean field (RMF) theory [1,2] has of late enjoyed
enormous success in providing an appropriate description of the
ground-state properties of nuclei. It has also been received
as a viable alternative to the phenomenological density dependent
Skyrme forces [3].  The built-in spin-orbit interaction
of the Dirac description of nucleons renders an attractive feature.
On the other hand, the spin-orbit interaction is added only
phenomenologically in the non-relativistic Skyrme forces.
In the RMF theory the saturation and the density dependence
of the nuclear interaction is provided by a balance between large
attractive scalar $\sigma$ meson and large repulsive vector $\omega$
meson. The asymmetry component is provided by the isovector $\rho$
meson. The nuclear interaction is hence generated by the exchange of
various mesons between nucleons. It is surmised that the structure of
the force in the RMF theory and its density dependence thus differ from
that of Skyrme forces, where in the latter a certain density dependence
is assumed at the outset.

There have been attempts [2] to understand the similarities and
differences between the RMF theory and the Skyrme approaches.
Although similarities are found, basic differences are, however, not
yet  understood. One of the
differences in these approaches is the acceptability of the
different values of the effective nucleon mass. It has been shown [4] that
for a satisfactory description of the spin-orbit splitting in nuclei,
a lower value of the effective mass around 0.55-0.60 is required
in the RMF theory. This is in contrast to the effective mass about
0.80 obtained from the Skyrme forces. A larger value of the
effective mass in the RMF theory would produce a very small spin-orbit
splitting, that would be incompatible with the observed magnitude.
This effect was also demonstrated [5] clearly in other variants of the ansatz
for scalar coupling, whereby the couplings leading to higher effective
mass showed an inadequacy in describing the spin-orbit splitting.
This difference in the requirement of the effective masses in the Skyrme
and RMF approaches is incomprehensible. It has, however, been pointed
out [6] that the two effective masses should not be compared, as the
two values do not correspond to the same property.

The charge radii of Pb isotopes and their isotope shifts have been a
matter of detailed discussion [7] within the framework of the Skyrme
forces. The isotopic chain of Pb nuclei exhibit a well known kink
in the behaviour of the empirical isotope shifts [8]. This implies that with
a gradual addition of neutrons, the charge radii of nuclei heavier than
$^{208}$Pb do not show a trend similar to that of lighter isotopes.
The Skyrme forces do succeed in describing the isotopes shifts
and thus charge radii of nuclei on the lighter side of $^{208}$Pb, where
a density-dependent pairing force is required to be assumed.
All the Skyrme forces, however, fail to reproduce
the isotope shifts of the heavier counterparts, as has been discussed
in detail in Ref. 7.  This indicates that remaining within the Skyrme
ansatz, the mean field is not able to describe heavier isotopes of Pb.
In the present work, we investigate this long standing problem within the
framework of the RMF theory.

As stated above, the RMF theory represents a framework that provides
an excellent description of the ground-state properties of nuclei [9].
The ansatz of the interaction in the RMF theory is based upon
Lagrangian of the form [1]:

\begin{equation}
\begin{array}{rl}
{\cal L} &=
\bar \psi (i\rlap{/}\partial -M) \psi +
\,{1\over2}\partial_\mu\sigma\partial^\mu\sigma - U(\sigma)
-{1\over 4} \Omega_{\mu\nu}\Omega^{\mu\nu} + \\
\                                           \\
\  & {1\over2}m_\omega^2\omega_\mu\omega^\mu
  -{1\over4}{\vec R}_{\mu\nu}{\vec R}^{\mu\nu} +
    {1\over2}m_\rho^2\vec\rho_\mu\vec\rho^\mu
     -{1\over 4}F_{\mu\nu}F^{\mu\nu} \\
\                                     \\
\  & -g_\sigma \bar\psi \sigma \psi~
    -~g_\omega \bar\psi \rlap{/}\omega \psi~
    -~g_\rho   \bar\psi \rlap{/}\vec\rho\vec\tau\psi
    -~e        \bar\psi A\llap{/}\psi.
\end{array}
\end{equation}
\medskip
\par\noindent
where the Dirac nucleon interacts with the $\sigma$ and $\omega$ meson
fields. The $\rho$ meson generates the isovector component of the force.
The nonlinear $\sigma\omega\rho$ model which we employ, has a nonlinear
scalar self-interaction of the $\sigma$ mesons as given by

\begin{equation}
U(\sigma)~=~{1\over 2}m_\sigma^2\sigma^2~+~
{1\over 3}g_2\sigma^3~+~{1\over 4}g_3\sigma^4,
\end{equation}

where $g_2$ and $g_3$ are the non-linear parameters. Details on the
RMF theory have been discussed in Ref. 2. There have been a few parameter
sets [2] in the RMF theory, which have been used extensively to obtain the
ground-state properties, the sets NL1 [10] and NL2 [11] being among the
few. Both the above forces provide neutron radii and thus neutron skin
thickness [12] of neutron-rich nuclei much larger than the empirical values.
Recently, the ground-state properties of nuclei have been investigated [9]
 in the non-linear $\sigma\omega\rho$ model. It has been noted that indeed
a stronger $\rho$ meson coupling and hence a very large asymmetry
energy of the above forces gives a larger neutron skin thickness of
neutron-rich nuclei. Consequently, a new force NL-SH has been obtained,
where the above defect of the earlier forces has been removed. It has also
been shown [9] that this force describes the ground-state binding energies,
charge and neutron radii of spherical as well as of deformed nuclei very far
off the stability line very well.  We employ the above force to obtain the
ground-state properties of the isotopic chain of Pb nuclei. The parameters
of the force NL-SH are the following:

\par\noindent
M = 939.0 MeV; $m_\sigma$ = 526.059 MeV; $m_\omega$ = 783.0 MeV; $m_\rho$ =
763.0 MeV;
\par\noindent
$g_\sigma$ = 10.444; $g_\omega$ = 12.945; $g_\rho$ = 4.383; $g_2$ = $-$6.9099;
$g_3$ = $-$15.8337.

The calculations have been performed within the Hartree approximation.
Although most of the Pb isotopes close to $^{208}$Pb are spherical,
an axially symmetric configuration has been assumed and Hartree
minimization has been performed. The method of the oscillator expansion [13]
has been employed, whereby both the fermionic as well as bosonic
wavefunctions have been expanded in N = 12 shells. We have considered
all the even mass Pb isotopes from A = 190 to 214. For convergence reasons,
N = 14 Fermionic shells have also been considered. It has been found
that the difference between the N=12 and N=14 calculations is miniscule.
Thus, we present only the results obtained with N=12.
For all the open-shell nuclei pairing has been included within the BCS
formalism with the pairing gap obtained from the particle separation
energies of the neighbouring nuclei. The quadrupole deformations obtained
from the convergence are very small and correspond essentially to spherical
configuration for all the nuclei considered here.
\medskip
Figure 1 shows the binding energy per nucleon of Pb isotopes obtained in
the relativistic Hartree approximation. The empirical binding energies are also
shown for comparison. It can be seen clearly that the binding energies of all
isotopes are reproduced well for the set NL-SH. The deviations are at most
0.1\%, and  in many cases better. The overall agreement with the empirical data
is remarkable. The calculated energies for the set NL1 show agreement with the
empirical binding energies only in the vicinity of $^{208}$Pb.
This is expected as the force NL1 was obtained [10] by fitting binding
energies including those of $^{208}$Pb. In going to the lighter side of the
isotopic chain, the binding energies obtained with NL1 show a systematic
deviation from the empirical data. The difference between the calculated
and the empirical binding energies increases as the neutron excess decreases.
This discrepancy is inevitably due to the large asymmetry energy of about
44 MeV for NL1. The problem has been rectified [9] in the force NL-SH.

The rms charge radii have been obtained as usual by folding
the proton form factor with the proton density distribution.
The charge radii of Pb isotopes  obtained for NL-SH and NL1 are
shown in Fig. 2. The empirical charge radii for Pb isotopes, as shown
in this figure, have been obtained from the measured isotope shifts [8]
employing the empirical [14] charge radius of $^{208}$Pb as 5.503 fm.
The charge radii from NL-SH follow closely the empirical ones implying
that the experimental data is well described by NL-SH. The force NL1,
on the other hand, overestimates the charge radii of Pb isotopes and
especially those of the lighter neutron-deficient ones.

The calculated charge radii have been used to obtain the isotope
shifts. The nucleus $^{208}$Pb has been taken as the reference point.
As in ref. 7, the isotope shifts ($\Delta r_c^2 = r_c^2(A) - r_c^2(208)$) have
been modified by substracting an equivalent of the liquid-drop difference ($
\Delta r_{LD}^2 = r_{LD}^2(A) - r_{LD}^2(208)$) obtained from $r^2_{LD}(A) =
{3\over 5}r_0^2A^{2/3}$, where $r_0$ = 1.2 fm. Fig. 3 shows the calculated
value of the modified isotope shifts ($\Delta r_c^2 - \Delta r_{LD}^2$)
for Pb nuclei. All the data have been presented in the same fashion.
The empirical values are from the precision data obtained from the atomic beam
laser spectroscopy [8]. The empirical data exhibit
a conspicuous kink about $^{208}$Pb. The figure also shows the
theoretically obtained isotope shifts for the two forces NL-SH and
NL1 and compare them with those from Skyrme interaction SkM*. Remarkably,
isotope shifts from NL-SH reproduce the kink
very well. Only below A = 198 the theoretical isotope shifts
show a systematic divergence from the empirical data. This behaviour
is well below the kink and may be attributed to the transitional
behaviour of nuclei in the light Pb isotopes. This region of mass number
is prone to such effects. The force NL1, on the other hand,
shows a reasonable kink on the higher side of $^{208}$Pb.
On the lower side, NL1 shows a slight divergence from the data and
the slope of the theoretical values is also different from
that of the empirical data. It may be noted that due to inaccurate
description of the other ground-state properties and a very large
asymmetry energy, NL1 is not expected to provide a proper
behaviour of the isotope shifts. In Fig. 3 we also show for comparison
isotope shifts for SkM* as taken from ref. 7. The data points
on $^{194}$Pb and $^{214}$Pb are representative of the behaviour of SKM*
on the isotope shifts. On the lighter side, SkM* shows a behaviour similar
to NL-SH. It, however, shows an almost linear function with mass number and
consequently displays a conspicuous divergence from the empirical data on
heavier side.

The kink in the experimental data implies that some changes
in placement of extra neutrons in a new shell take place, which
bring about this kink. Attempts to reproduce this kink within
the density-dependent Skyrme forces have been discussed
in detail in Ref. 7. As shown in fig. 2 of Ref. 7, within the
Hartree-Fock mean field, all the Skyrme forces e.g. SkM*, Ska and SIII
show a strong deviation from the empirical data in the isotope
shifts for nuclei heavier than $^{208}$Pb. The lighter Pb isotopes could,
however, be described by SkM* and SGII. The binding energies, on the
other hand, show a behaviour opposite to that of isotope shifts.
Whereas SkM* reproduces the binding energies of isotopes including and
heavier than $^{208}$Pb, it shows a systematic divergence from empirical
binding energies for lighter isotopes, the difference being accentuated on
going to the neutron-deficient side. The other two Skyrme forces show
disagreements with the empirical binding energies on both the sides.
Taking all possible corrections beyond the mean field into account does not
resolve the discrepancy. It has been demonstrated unambiguously in Ref. 7
that Skyrme forces face a tremendous problem in describing the binding
energies and isotopes shifts of nuclei away from $^{208}$Pb. It has also
been concluded [7] that the compressibility of the nuclear matter does
not have a decisive effect on the isotope shifts of nuclei in contrast
to what was stated earlier in ref. [15]. It is worth
noting that most of the Skyrme forces have been fitted to
the binding energies and charge radii of spherical nuclei from $^{16}$O
to $^{208}$Pb, with almost comparable surface and asymmetry energies,
but with a strong difference in the incompressibility and the equation
of state (EOS). It has been shown [16] that even with a very broad
variation in the EOS, one can construct any number of Skyrme interactions
that could describe the ground-state properties of some spherical nuclei
from $^{16}$O to $^{208}$Pb. The forces SkM*, SGII and SIII manifest
a similar scenerio as discussed in Ref. [16]. The failure of these
well-known Skyrme forces to reproduce the empirical data on Pb
isotopes raises an important question as to whether the Skyrme ansatz could
be extrapolated to describe nuclei far off the stability notwithstanding
the problem associated with the Skyrme forces even on the stability line.

The ability to reproduce the kink by the RMF theory and the
failure of the present Skyrme interactions not to be able to do so,
raises some important issues. The basic shell effects that
are inherent in the structure of nuclei over the periodic
table provide a clue as to how effectively various
theories are able to take these effects into account. The kink
in the empirical data on charge radii of Pb isotopes belongs
to these questions. Evidentally, the RMF theory succeeds in
accommodating these shell effects which run across the
shell-closure. The shell effects across the magic numbers
play a significant role in astrophysical r-processes.

The behaviour of the Skyrme mean-field approach and of
the RMF theory towards isotope shifts indicates an important
difference between the two approaches. This difference is
not easy to explain. However, examining the two approaches
we feel that the basic density dependence of the two
interactions and the difference in the saturation mechanism
of the two methods could be at the origin of the difference
in the isotope shifts. Another important aspect that is
different is the spin-orbit interaction. Whereas in the RMF
theory, the spin-orbit interaction orginates from the
coupling of $\sigma$ and $\omega$ mesons to Dirac nucleon,
the spin-orbit term in the Skyrme approach is added
phenomenologically. The spin-orbit splitting is responsible for
putting different orbitals in space and thus determining the
structure of a nucleus. Thus, a difference in the spin-orbit
splitting in the two methods would contribute to the
difference in the isotope shifts of the two approaches.
Further work in this direction is in progress.
\medskip
In conclusion, we have examined the behaviour of the RMF theory
in describing the anomalous isotope shifts in Pb nuclei.
It has been shown that the RMF theory describes the isotope shifts
successfully in marked constrast to the Skyrme mean field approach.
This difference has unveiled some issues underlying
the applicability of the two approaches in nuclear structure.
It is expected that these differences will have broader implications
on the properties of neutron-rich nuclei far from the stability line, and in
particular, for nuclei close to neutron-drip line, with
consequences on the EOS of neutron matter and neutron-star structure.

\bigskip
This work is supported in part by the Bundesministerium f\"ur Forschung und
Technologie. One of the authors (G.A.L.) acknowledges support from the Deutsche
Akademische Austauschdienst (DAAD). M.M.S. would like to thank Ben Mottelson
and David Brink for their kind hospitality at ECT$^*$, Trento.

\vfill
\eject

\leftline{\bf Figure Captions.}
\bigskip
\bigskip
{\bf Fig. 1} The binding energies of Pb isotopes obtained in the
relativistic Hartree approximation with the forces NL1 and NL-SH. The
empirical values (expt.) are also shown for comparison.
\bigskip

{\bf Fig. 2} The charge radii of Pb isotopes obtained with
NL1 and NL-SH. The empirical values (expt.) from laser spectroscopic
 measurements [9] follow closely the charge radii from NL-SH.
\bigskip

{\bf Fig. 3.} The isotope shifts of Pb nuclei obtained with NL1 and NL-SH.
The empirical values [9] along with the values from NL-SH exhibit a
conspicuous kink in the isotopes shifts about $^{208}$Pb. The SkM* values [7]
show a large deviation from the empirical data for heavier nuclei.

\bigskip
\vfill

\end{document}